\documentclass[lettersize,journal]{IEEEtran}
\usepackage{amsmath,amsfonts}
\usepackage{algorithmic}
\usepackage{algorithm}
\usepackage{array}
\usepackage[caption=false,font=normalsize,labelfont=sf,textfont=sf]{subfig}
\usepackage{textcomp}
\usepackage{stfloats}
\usepackage{url}
\usepackage{verbatim}
\usepackage{graphicx}
\usepackage{cite}
\usepackage{multirow}
\usepackage{booktabs} 
\usepackage{xcolor}
\usepackage{colortbl}
\usepackage{hyperref}

\begin{document}

\title{FCA²: Frame Compression-Aware Autoencoder for Modular and Fast Compressed Video Super-Resolution}

\author{Zhaoyang Wang, Jie Li, Wen Lu, \textit{Member, IEEE,} Lihuo He, \textit{Member, IEEE,} Maoguo Gong, \textit{Fellow, IEEE,} Xinbo Gao, \textit{Fellow, IEEE} 
\thanks{
This work was supported in part by the National Natural Science Foundation of China under Grants 62036007, 62101084, 62221005, 62171340, 62476207; in part by the Chongqing Natural Science Foundation Innovation and Development Joint Fund Project under Grants CSTB2023NSCQ-LZX0085 and CSTB2023NSCQ-BHX0187; in part by the Key Industrial Innovation Chain Project in Industrial Domain of Shaanxi Province under Grant No. 2020ZDLGY05-01; in part by the Fundamental Research Funds for the Central Universities under Grant No.  YJSJ25008. (Corresponding authors: Xinbo Gao.)

Zhaoyang Wang, Jie Li, Wen Lu and Lihuo He are with the State Key Laboratory of Integrated Services Networks, School of Electronic Engineering, Xidian University, Xi’an, Shaanxi 710071, China (e-mail: zywang23@stu.xidian.edu.cn; leejie@mail.xidian.edu.cn, luwen@mail.xidian.edu.cn and lhhe@mail.xidian.edu.cn). 

Maoguo Gong is with the Key Laboratory of Collaborative Intelligence Systems, Ministry of Education, Xidian University, Xi’an 710071, China, and is also affiliated with the College of Mathematical Science, Inner Mongolia Normal University, Hohhot 010028, China (e-mail: gong@ieee.org).

Xinbo Gao is with the School of Electronic Engineering, Xidian University, Xi’an 710071, China (e-mail: xbgao@mail.xidian.edu.cn), and with the Chongqing Key Laboratory of Image Cognition, Chongqing University of Posts and Telecommunications, Chongqing 400065, China (e-mail: gaoxb@cqupt.edu.cn).
}}

\markboth{Journal of \LaTeX\ Class Files,~Vol.~14, No.~8, August~2021}%
{Shell \MakeLowercase{\textit{et al.}}: A Sample Article Using IEEEtran.cls for IEEE Journals}


\maketitle

\begin{abstract}
State-of-the-art (SOTA) compressed video super-resolution (CVSR) models face persistent challenges, including prolonged inference time, complex training pipelines, and reliance on auxiliary information. As video frame rates continue to increase, the diminishing inter-frame differences further expose the limitations of traditional frame-to-frame information exploitation methods, which are inadequate for addressing current video super-resolution (VSR) demands. To overcome these challenges, we propose an efficient and scalable solution inspired by the structural and statistical similarities between hyperspectral images (HSI) and video data. Our approach introduces a compression-driven dimensionality reduction strategy that reduces computational complexity, accelerates inference, and enhances the extraction of temporal information across frames. The proposed modular architecture is designed for seamless integration with existing VSR frameworks, ensuring strong adaptability and transferability across diverse applications. Experimental results demonstrate that our method achieves performance on par with, or surpassing, the current SOTA models, while significantly reducing inference time. By addressing key bottlenecks in CVSR, our work offers a practical and efficient pathway for advancing VSR technology.
Our code will be publicly available at \url{https://github.com/handsomewzy/FCA2}.
\end{abstract}


\begin{figure}[t]
    \centering
    \includegraphics[width=0.9\linewidth]{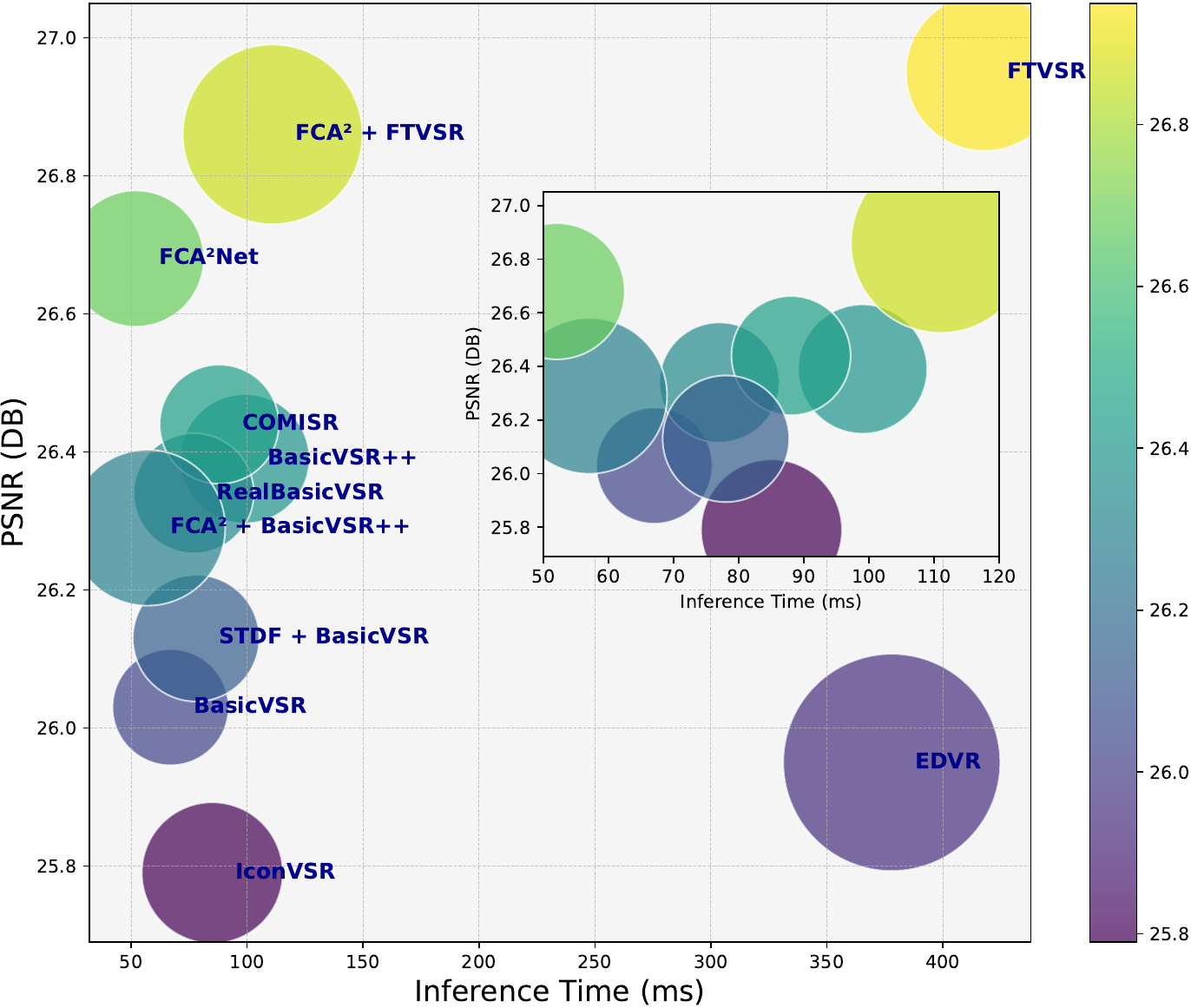}
    \caption{Visualization of the trade-off between inference time (x-axis) and PSNR (y-axis), with circle size representing the model size. Larger circles indicate larger models, highlighting the balance between performance and computational efficiency.}
    \label{fig1}
    \vspace{-3mm}
\end{figure}

\section{Introduction}
\IEEEPARstart{T}{he} primary objective of video super-resolution (VSR) is to reconstruct a high-resolution (HR) video sequence from its low-resolution  (LR) counterpart. With the advent of deep learning, numerous advanced methods have been proposed \cite{fu2025global, xiao2023online, wang2018multi, baniya2023omnidirectional,zhu2022fffn,chan2021basicvsr,chan2022basicvsr++,chan2022investigating, fuoli2019efficient, isobe2022look}, yielding impressive results. 
VSR methods based on recurrent architectures, which leverage both preceding and subsequent frames to maximize the utilization of video sequence information, are increasingly becoming mainstream \cite{chan2021basicvsr,chan2022basicvsr++,chan2022investigating}.
However, many of these approaches are trained under idealized conditions, typically assuming a single type of distortion, such as bilinear interpolation downsampling. Models trained in this manner often struggle to generalize to real-world video data, where distortions are more complex and diverse \cite{chan2022investigating}.

To address this challenge, several blind super-resolution (SR) models \cite{pan2021deep,lee2021dynavsr, chan2022investigating, qiu2023dual,huang2020fast,chen2023dynamic} have been proposed to handle more realistic scenarios where the specific type of distortion is unknown. However, in practical applications, most video data is captured and stored by digital devices or hosted on online platforms, where compression strategies are applied at varying levels \cite{isobe2020intra,isobe2021multi,rippel2019learned,wang2023compression}. For instance, videos on streaming platforms may include content in formats ranging from standard definition (SD) to ultra-high definition (Ultra HD), each subject to different levels of compression. While the distortion introduced by such compression is not entirely unknown, it differs from the traditional VSR task, as well as from fully blind SR tasks. This scenario is characterized as compressed VSR (CVSR) \cite{li2021comisr,wang2023compression,qiu2022learning}.

When the compression is shallow, traditional VSR methods, as well as certain blind VSR approaches, exhibit negligible performance degradation. However, as the compression level increases significantly, or varies across different levels, these methods tend to show suboptimal performance \cite{wang2023compression}.
To address this issue, a straightforward approach \cite{deng2020spatio,guan2019mfqe,xu2022transcoded,yang2018multi} is to first decompress the data to mitigate the impact of varying compression levels on the video content, and subsequently integrate it into the conventional VSR model for SR.
However, when confronted with varying compression levels, the model struggles to differentiate between them and fails to adaptively estimate the compression degree, resulting in unsatisfactory performance \cite{li2021comisr,wang2023compression,qiu2022learning}.

\begin{figure*}[t]
    \centering
    \includegraphics[width=0.8\linewidth]{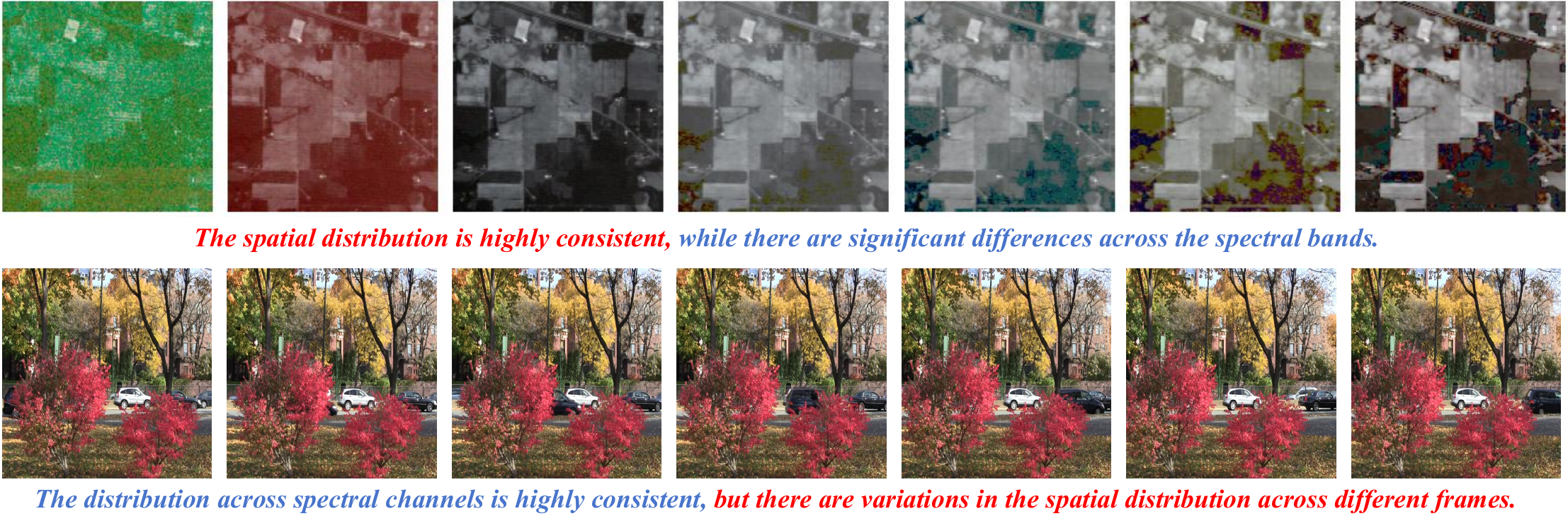}
    \caption{
Comparison of features between video data and hyperspectral data. The first row shows pseudocolor illustrations of three selected bands from the IndianPines \cite{song2024interactive} dataset, while the second row displays the Vid4 \cite{liu2013bayesian} dataset.}
    \label{hsi_video}
\end{figure*}

\begin{figure*}[t]
    \centering
    \includegraphics[width=0.8\linewidth]{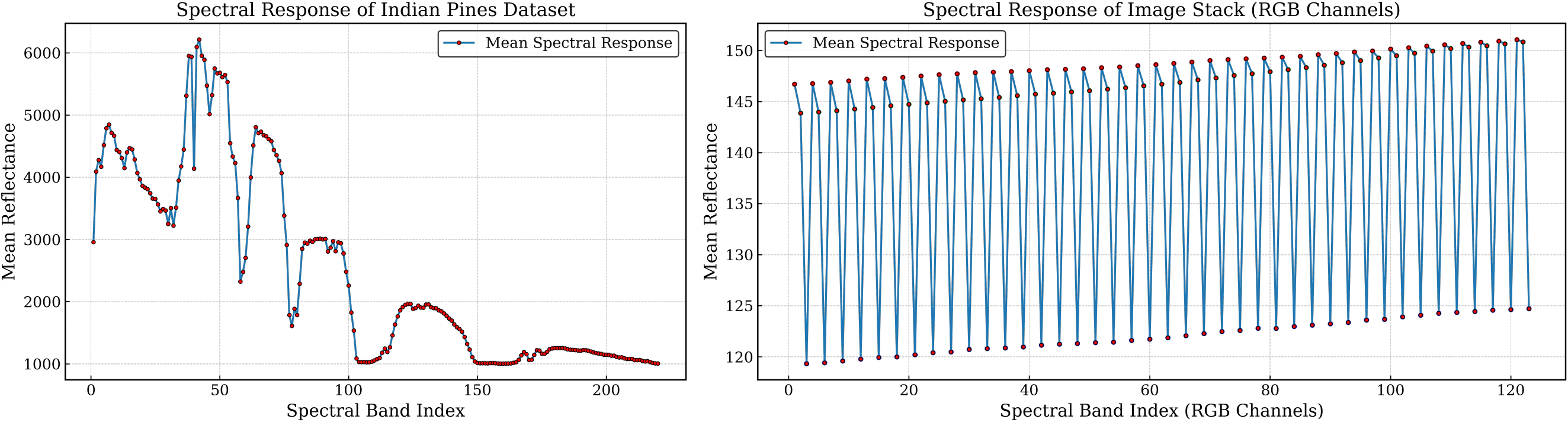}
    \caption{
Comparison of channel values between HSI data and video data stacked by frame. The analysis is conducted on the IndianPines \cite{song2024interactive} dataset for HSI data and the Calendar video from the Vid4 \cite{liu2013bayesian} dataset for video data. HSI exhibits significant spectral variability across bands, whereas video data maintains a consistent RGB distribution pattern.}
    \label{hsi_video_compare}
    \vspace{-3mm}
\end{figure*}

Recently, several pioneering works have been proposed to tackle the CVSR problem. In \cite{li2021comisr}, Li \textit{et al.} leverage optical flow features and a Laplace enhancement module to address variations induced by different compression levels. In \cite{qiu2022learning}, a Transformer-based approach is employed, incorporating frequency domain features to mitigate the effects of varying compression levels. Meanwhile, in \cite{wang2023compression, chen2024multi, zhu2024deep}, the authors additionally train a compression level encoder to extract features corresponding to different compression levels, which are then used as conditional information to guide the VSR network. This process requires the additional compression feature information as a guide to effectively adapt to varying compression levels. While these methods have demonstrated promising results, several challenges remain:
\begin{itemize}
\item \textbf{Limited utilization of inter-frame information:} Most VSR networks rely heavily on the preceding and succeeding frames, but the utilization of inter-frame information remains insufficient. As video frame rates continue to increase, the differences between consecutive frames become smaller, making the exclusive use of only the front and back frames an inefficient way to capture the inter-frame information.
\item \textbf{Slow inference speed:} While Transformer-based models significantly enhance performance, they struggle to meet the practical demands of real-time inference due to their slow processing speed \cite{qiu2022learning}.
\item \textbf{Additional metadata information:} To address the CVSR problem, some models rely on additional compression feature information as higher-level guidance \cite{wang2023compression} (e.g., variables from the encoding process in H.264 \cite{wiegand2003overview}). However, such intermediate variables are often not explicitly available or easily accessible in practical applications.
\end{itemize}

Additionally, rapid advancements in digital imaging technologies have resulted in higher video frame rates, consequently reducing temporal variations between adjacent frames. Under such conditions, traditional methods that heavily rely on inter-frame differences become increasingly inefficient, highlighting the necessity for novel approaches to effectively handle redundant frame information and improve reconstruction quality. 

This scenario closely parallels the rich channel information inherent in remote sensing imagery, as illustrated in Fig. \ref{hsi_video} and Fig. \ref{hsi_video_compare}. When a video sequence is arranged as a three-dimensional (3D) data cube by stacking individual frames, its structure becomes highly analogous to hyperspectral images (HSI) data commonly encountered in remote sensing. Both data modalities exhibit similar characteristics in feature distribution, particularly in terms of redundancy and inter-frame (or spectral band) correlations.
The spatial structure of HSI data is \textit{\textbf{consistent across channels but exhibits notable spectral differences,}} closely resembling video sequences, \textit{\textbf{where frames maintain similar RGB channel distributions yet show subtle spatial variations}} (see Fig. \ref{hsi_video_compare} and Fig. \ref{hsi_video}). 
Motivated by this similarity, we explore the potential application of HSI-SR processing techniques to CVSR tasks. Specifically, to address the redundancy across spectral channels in HSI data, channel compression methods have been widely adopted \cite{jiang2020learning, wang2024enhancing, wang2022group}. Drawing inspiration from these approaches, we propose a novel frame compression module tailored to CVSR, which effectively reduces frame redundancy and accelerates multi-frame inference.

To address the aforementioned challenges, we propose the \textbf{F}rame \textbf{C}ompression-\textbf{A}ware \textbf{A}utoencoder (FCA²) for the CVSR problem. Our module is highly adaptable and effectively resolves the issue of under-utilization of inter-frame information, while significantly accelerating inference speed. Specifically, we design a frame compression module tailored to different compression levels. Given the improvements in video quality, inter-frame differences have become smaller, and multiple neighboring frames are compressed into a single frame, thereby eliminating redundant information. The proposed process condenses inter-frame dependencies and integrates multi-level compression cues to extract prior knowledge. By reducing the number of frames to be processed, the VSR model focuses only on key compressed frames, substantially accelerating inference (see Fig.~\ref{fig1}).
To enhance coding efficiency, we propose an optical flow-guided coding module that integrates both spatial and temporal coding-decoding strategies. Furthermore, we introduce a group-based compression mechanism, frame averaging techniques, and a lightweight color correction module to improve representation quality. The main contributions of this work are summarized as follows:

\begin{itemize}
    \item Inspired by the structural and statistical similarities between HSI and video data, we introduce a frame compression module to CVSR, enhancing inter-frame information utilization, accelerating inference, and ensuring seamless integration with existing VSR frameworks.
    \item We propose a frame-based grouping strategy, an optical flow-guided coding module, and a color correction module to achieve efficient frame compression and decompression.
    \item Extensive experiments demonstrate the effectiveness of the model, which not only accelerates inference speed but also enhances inter-frame information utilization, significantly improving overall performance. The results are comparable to state-of-the-art (SOTA) performance.
\end{itemize}

The remainder of this paper is organized as follows. Section II reviews related work. Section III describes the proposed module in detail, including its design principles and implementation. Experimental results and analyses are presented in Section IV. 
In Section V, we outline the limitations of the proposed
model, potential avenues for improvement, and our future
work directions.
Finally, Section VI concludes the paper.

\section{Related Work}
\subsection{Video Super-Resolution}
Current VSR methods can be broadly classified into two categories based on how they handle time dimension data: sliding-window-based methods and recurrent-based methods. Sliding-window-based approaches \cite{tao2017detail,xue2019video ,caballero2017real} typically process a set of frames and traverse them in their entirety, during which optical flow is computed to align inter-frame differences. Alternatively, deformable convolutions \cite{dai2017deformable,wang2019edvr,tian2020tdan} are used to align features across frames. Some methods \cite{isobe2020video,li2020mucan, yi2019progressive} also leverage motion compensation mechanisms based on 3D convolutional neural networks for improved feature alignment.
These methods are still limited in their utilization of inter-frame information and struggle to capture long-range dependencies across frames. Currently, the dominant VSR approaches are recurrent-based methods, which can be further divided into unidirectional and bidirectional methods. Unidirectional methods \cite{fuoli2019efficient,sajjadi2018frame, isobe2020video1} are primarily employed for real-time processing, utilizing only past video sequences as feature information and operating in an online fashion. In contrast, bidirectional methods \cite{chan2021basicvsr,chan2022basicvsr++,chan2022investigating,li2023multi,haris2019recurrent} use both past and future frames for SR, leading to improved results.

In \cite{chan2021basicvsr}, Chan \textit{et al.} leverage both forward and reverse feature information based on an RNN architecture to construct the BasicVSR network, which serves as the backbone for most subsequent VSR networks. Building on this, Chan \textit{et al.} further improve inter-frame utilization by proposing BasicVSR++ \cite{chan2022basicvsr++}, which extends the contribution of information from both the past and future frames. Additionally, BasicVSR++ performs feature transfer directly in the subsequent feature layers, thereby further enhancing the utilization of inter-frame information.
In \cite{li2023multi}, building upon the BasicVSR architecture, Li \textit{et al.} introduce the multi-frequency representation enhancement module to further improve inter-frame information interaction.
To further enhance information mining capabilities, in \cite{zhou2024video}, Zhou \textit{et al.} introduce a mask operation with inter-intra-frame attention to improve inter-frame utilization. Additionally, Zhou \textit{et al.} have leveraged the diffusion model to design VSR networks \cite{zhou2024upscale}, harnessing its powerful feature extraction capabilities to mine inter-frame information.

Most existing methods aim to enhance the model's capacity to effectively utilize and exploit inter-frame information. However, these approaches often make idealized assumptions, such as generating LR samples via bicubic interpolation. This simplification fails to capture the complexities inherent in the CVSR task, where such methods struggle to handle compression artifacts and real-world distortions effectively.

\subsection{Compressed Video Super-Resolution}
Several successful attempts have been made on CVSR tasks. MFQE \cite{guan2019mfqe,yang2018multi} initially proposed a lightweight multi-frame CNN, leveraging peak-quality reference frames identified through quality assessment, to improve the visual quality of surrounding degraded frames. The paper also highlights the limitations of single-frame SR methods \cite{yang2018enhancing,yang2017decoder}, suggesting that the enhanced exploitation of inter-frame information has become the prevailing approach in the field.
In \cite{deng2020spatio}, Deng \textit{et al.} propose spatial-temporal deformable convolution to aggregate optical flow information, effectively mitigating perturbations at various compression levels.
In \cite{chan2022investigating}, Chan \textit{et al.} propose a carefully designed data-noising model, which artificially generates various types of LR images. They pre-train a cleaning network and fit these LR images to a common distribution, effectively addressing the CVSR problem.

In addition to the direct decompression approach, several methods have been proposed that incorporate frequency domain information or leverage more advanced models to tackle the CVSR problem. In \cite{li2021comisr}, Li \textit{et al.} design a detail-aware flow estimation module and introduce a Laplace enhancement module to better integrate optical flow information and incorporate high-frequency feature details. In \cite{qiu2022learning}, Qiu \textit{et al.} map the image into the frequency domain and develop a frequency domain Transformer combined with a self-attention mechanism, enabling spatial and temporal collaboration for the restoration of high-frequency feature information.

Subsequently, researchers have shifted focus towards utilizing additional metadata generated during the CVSR compression process as auxiliary information to guide network training. In \cite{wang2023compression}, Wang \textit{et al.} leverage intermediate coded metadata produced by H.264 compression to assist in network training. In \cite{zhu2024deep}, Zhu \textit{et al.} leverage compression-related priors, including partition maps, residual data, predicted frames, and motion vectors, as auxiliary cues. While the inclusion of these additional metadata improves model performance, their availability is often limited in practical applications, posing challenges for real-world deployment.



\subsection{Related Work in Video Compression and Coding}
Video encoding and compression are widely utilized to achieve efficient storage, faster transmission, and reduced computational resource consumption \cite{girod2005distributed, liu2020deep, duan2020video, sheng2022temporal, li2024perceptual}. Beyond these traditional purposes, several studies have also leveraged video coding techniques for inter-frame prediction \cite{li2024uniformly, li2024object} in downstream vision tasks. For instance, in \cite{li2024object} explores the use of coding information to enhance target detection performance. Moreover, concepts similar to group-wise coding strategies have been explored in related domains, such as temporal wavelet-based low-complexity quality enhancement \cite{dong2023temporal} and space-time video super-resolution \cite{xiao2023space,xiao2021space}, demonstrating the effectiveness of structured temporal modeling and emphasizing the potential of group-based compression paradigms in advanced video processing tasks.

\section{Proposed Frame Compression-Aware Autoencoder }

\subsection{Overall Architecture}
\begin{figure*}[t]
    \centering
    \includegraphics[width=0.9\linewidth]{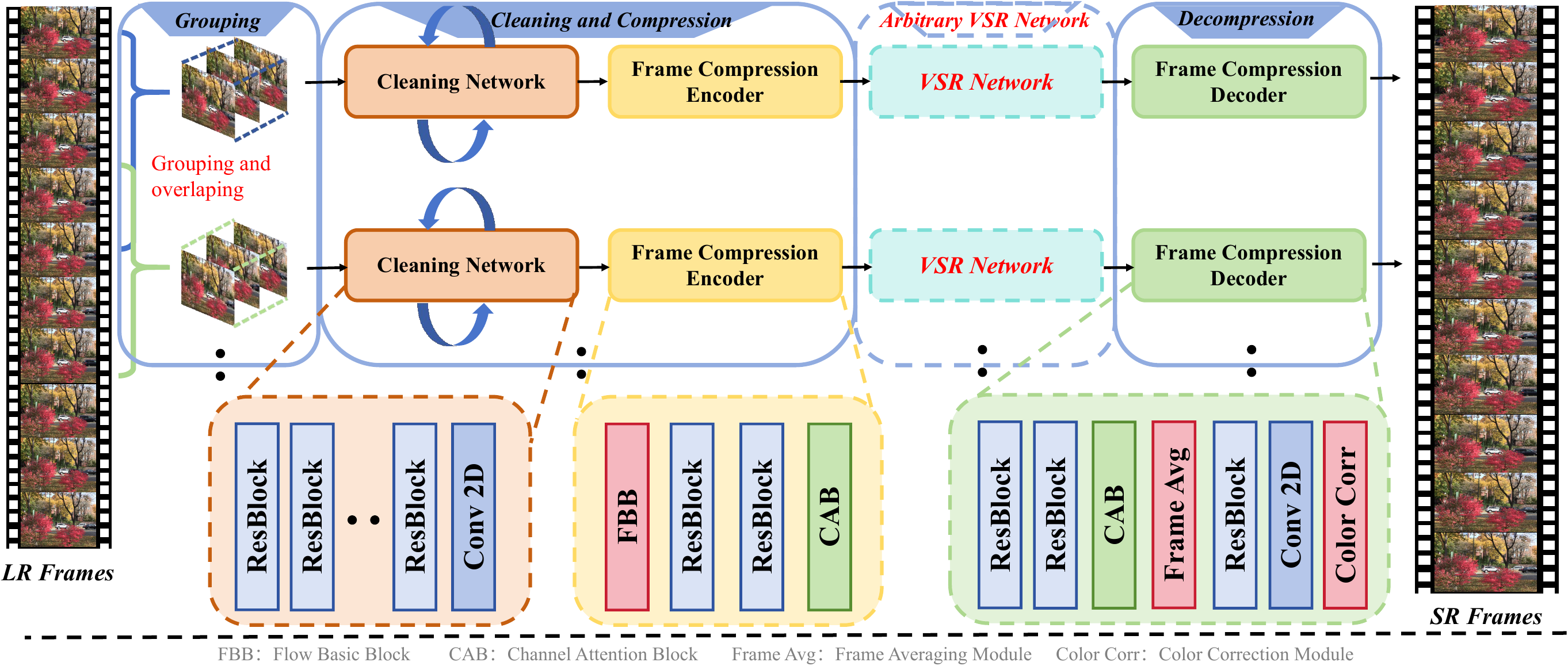}
    \caption{Illustration of the proposed FCA² model. The input video sequence initially undergoes overlapping grouping and encoding, followed by preliminary denoising applied independently to each group. Subsequently, frame compression encoding is performed to extract compact representations, which serve as input features for the subsequent VSR network. Finally, the outputs from the VSR network are upsampled, decoded, and concatenated to reconstruct the SR video sequence.}
    \label{main}
    \vspace{-3mm}
\end{figure*}

The overall architecture of our method is illustrated in Fig. \ref{main}. It comprises four key components: grouping, cleaning and frame compression, an arbitrary VSR network, and frame decompression. In the grouping stage, multi-frame video sequences are initially concatenated along the channel dimension to create 3D representations. A subset of channels is then selected for encoding with overlapping regions, thereby facilitating enhanced channel-wise interaction and more efficient information utilization.
The processed data is then passed to the cleaning and frame compression module, where simple distortions, such as pretzel noise and Gaussian noise, are mitigated. The cleaned data is subsequently fed into the compression encoder, which generates intermediate latent representations. These representations are input to an arbitrary VSR network to perform SR. The enhanced outputs are then sent to the frame decompression module, which reconstructs the original channel configuration used in the grouping stage. Finally, the decoded frames are reassembled and concatenated to form the super-resolved video sequence.

In the subsequent section, we elaborate on the key components and underlying design principles of each module in detail.

\subsection{Grouping Strategy}
Our grouping strategy is inspired by existing processing methods for HSI-SR. In \cite{jiang2020learning, wang2024enhancing, wang2022group}, group coding, over-segmentation, and compression are employed to reduce redundancy between HSI data channels, enhance information utilization, and minimize model parameters, thereby accelerating inference.
As illustrated in Fig. \ref{hsi_video}, the rapid advancement of digital technology has led to an increase in video frame rates, resulting in progressively smaller inter-frame differences. While HSI data exhibits significant variations across spectral channels but maintains spatial consistency, video data follows a different distribution pattern—channels adhere to RGB-based statistical properties (see Fig. \ref{hsi_video_compare}), while inter-frame variations occur only in localized spatial regions. Motivated by these observations, we design a grouping and compression strategy that effectively captures essential spatial-temporal dynamics, thereby reducing redundancy and facilitating efficient multi-frame inference.

To efficiently leverage inter-frame information, we first transform the multi-frame video sequence into a 3D representation similar to HSI data by concatenating frames along the channel dimension. Given an input sequence of \( N \) consecutive video frames:  

\begin{equation}
    \mathcal{X} = \{X_t\}_{t=1}^{N}, \quad X_t \in \mathbb{R}^{H \times W \times C}
    \label{eq1}
\end{equation}
where each frame \( X_t \) has spatial dimensions \( H \times W \) and \( C \) color channels (e.g., RGB with \( C=3 \)). We then apply a channel-wise concatenation operation to obtain a high-dimensional representation:
\begin{align}
    X_{\text{3D}} &= \text{Concat}(X_1, X_2, ..., X_N) \in \mathbb{R}^{H \times W \times (C \cdot N)} \nonumber \\
    &= [c_1, c_2, c_3, \dots, c_{C \cdot N}]
    \label{eq2}
\end{align}
where `Concat' represents concatenation along the channel axis, effectively expanding the feature dimension from \( C \) to \( C \cdot N \).  
This transformation allows the model to jointly process multiple frames as a single entity, improving the utilization of temporal information while preserving spatial consistency. Unlike traditional per-frame processing, this approach enhances feature interactions across neighboring frames, enabling more efficient redundancy reduction and feature compression in subsequent stages.

After concatenation, overlapping groups are constructed to enhance information interaction and utilization across channels, inspired by insights from previous studies \cite{jiang2020learning, wang2024enhancing, wang2022group}. This overlapping strategy ensures that adjacent groups share a portion of their information, facilitating better feature continuity and reducing the risk of information loss during compression. The final groupings are as follows:

\begin{equation}
    \begin{aligned}
        G &= [g_1, g_2, \dots, g_K] \\
          &= \underbrace{c_1, c_2, \dots, c_{S}}_{g_1}, 
          \underbrace{c_{S - O + 1}, \dots, c_{2S - O}}_{g_2}, \\
          &\underbrace{c_{2S - 2O + 1}, \dots, c_{3S - 2O}}_{g_3},  
          \dots,  \\
          &\underbrace{c_{(K-1)(S - O) + 1}, \dots, c_{K S - (K-1) O}}_{g_K}
    \end{aligned}
    \label{eq3}
\end{equation}
where \( S \) is the group size, representing the number of consecutive channels in each group, and \( O \) is the overlap factor, indicating the number of shared channels between adjacent groups. The index \( K \) denotes the total number of groups, ensuring complete coverage of the feature sequence. Each group \( g_k \) contains \( S \) channels, with an overlap of \( O \) channels with the previous group, facilitating better feature continuity and redundancy reduction. This structured overlapping mechanism enhances inter-group feature interaction, preventing abrupt discontinuities and improving downstream tasks.

\begin{figure*}[t]
    \centering
    \includegraphics[width=0.9\linewidth]{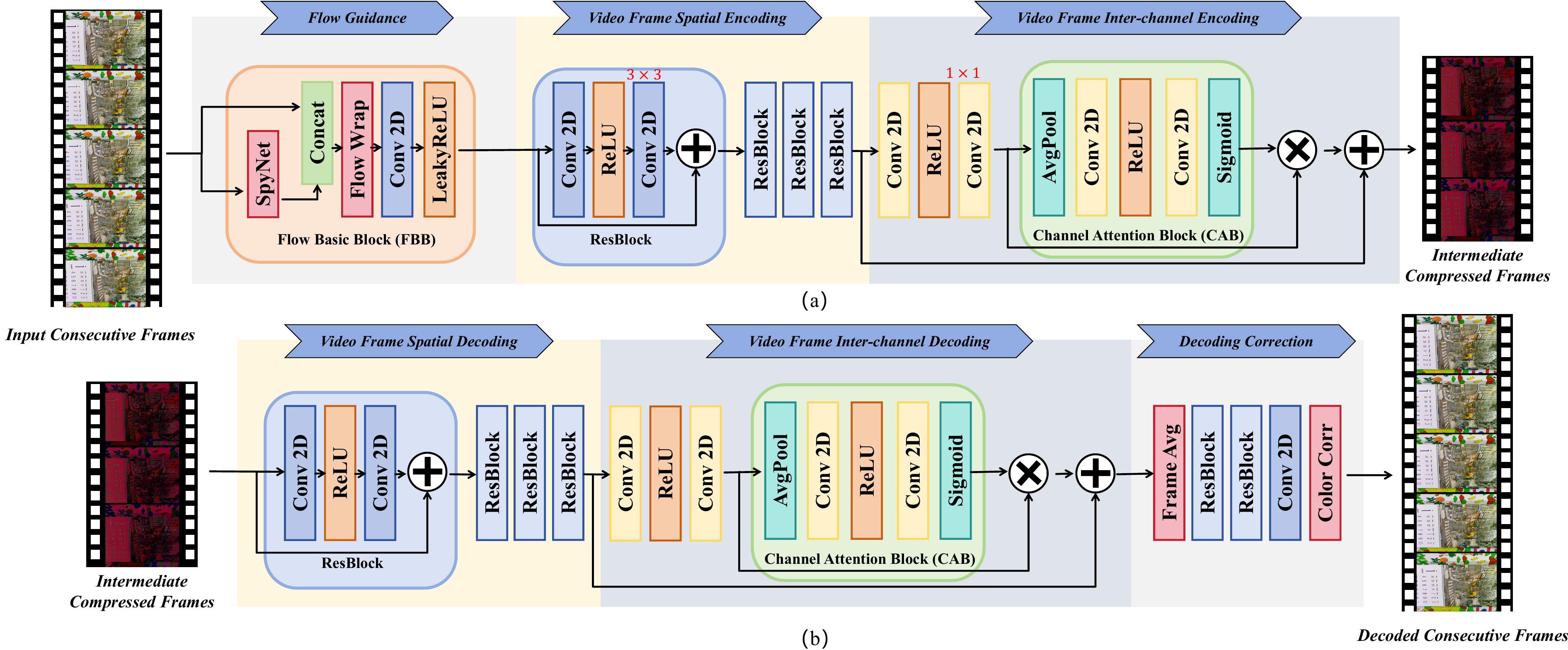}
    \caption{Illustrations of the encoder and decoder architectures. (a) depicts the encoder, and (b) illustrates the decoder. Both architectures are structured into three distinct stages to effectively process and reconstruct the video data.}
    \label{FCA²}
    \vspace{-3mm}
\end{figure*}

\subsection{Cleaning and Frame Compression Encoder}
\subsubsection{Cleaning Network}
Our cleaning network employs a straightforward architecture composed of several residual layers to perform basic noise reduction. In contrast to RealBasicVSR \cite{chan2022investigating}, which uses a dynamically updated cleaning module, our approach fixes the number of cleaning cycles to three, yielding faster inference without sacrificing performance. In \cite{chan2022investigating}, Chan \textit{et al.} propose a complex pre-training strategy that constructs training data with various distortion types to address the blind SR task. Here, we streamline this procedure by omitting the intricate pre-training steps and simplifying the training pipeline.
Specifically, we first pre-train the cleaning network on the existing PIPAL \cite{jinjin2020pipal} dataset, which contains a wide range of distortion types and thus negates the need for artificially generated noise. The resulting model parameters serve as the initial weights for our compression-aware task. We then conduct further pre-training on images encoded with different CRF values, encouraging these varying compression artifacts to converge to a common distortion space while simultaneously filtering out residual noise.

\subsubsection{Frame Compression Encoder}
As depicted in Fig. \ref{FCA²}, our Frame Compression Encoder comprises three main components: an optical-flow-guided encoding module, a spatially-oriented encoding module, and a channel-oriented encoding module leveraging channel attention. This architecture collectively exploits both motion and spatial cues, while incorporating channel-wise attention to enhance overall compression performance.

In the optical-flow-guided encoding module, we employ the same optical flow network used in previous work \cite{chan2021basicvsr, chan2022basicvsr++, wang2023compression}, adopting a similar conditional integration strategy. Specifically, we first predict the optical flow for neighboring frames and then concatenate these predictions into a single input for subsequent encoding. This optical flow information aids the model in capturing inter-frame dependencies and maintaining temporal consistency, while also focusing its attention on spatial regions with significant motion changes.

Next, the encoder performs spatial and channel coding. Specifically, a  \(3 \times 3\) convolutional kernel captures detailed spatial information, while a  \(1 \times 1\) kernel focuses on channel-specific encoding. To further enhance multi-channel temporal representations, we introduce a channel attention mechanism. The encoder's final output comprises compressed intermediate feature variables, which are then passed to the VSR network for subsequent processing.

\subsection{Frame Decompression Decoder}
The decoder comprises three main components, which can be seen in Fig. \ref{FCA²}. The first two components mirror those of the encoder, focusing on spatial decoding and channel decoding through an analogous architecture. As the optical flow information derived during encoding no longer retains a clear semantic meaning in the intermediate feature frames, no additional optical flow guidance is incorporated in the decoding stage.

Nonetheless, certain encoding operations—such as overlapping grouping, which may introduce repeated encoding, and multi-channel grouping, which can yield atypical color-channel distributions—necessitate further correction. To address these issues, the third component, a decoding correction module, is introduced. This module integrates a frame-averaging mechanism and a color correction procedure to refine and restore the output to its proper distribution.

\subsubsection{Frame-averaging Mechanism}
The overlapping introduced by the grouping procedure can cause certain channels to be encoded multiple times, potentially leading to redundancy and artifacts in the final output. To address this issue, we record the frequency of encoding for each channel, enabling an averaging operation during decoding. This averaging step refines the reassembled frames by mitigating the adverse effects of repeated encoding and ensuring consistency in the restored data.

\subsubsection{Color Correction}
In the encoding stage, we incorporate optical flow information between frames, thereby expanding the number of channels. When multiple channels are selected for encoding, these channels do not strictly adhere to the RGB color distribution, potentially introducing color distortion in the final decoded output. To address this issue, we introduce a color correction module that aligns the mean and variance of the final decoded output with those of the normal LR frame at the same time instant. This process can be expressed as:

\begin{equation}
\label{color_correction}
\mathrm{corrected\_output} = 
\left( \frac{\mathrm{sr\_output} - \boldsymbol{\mu}_{sr}}{\boldsymbol{\sigma}_{sr}} \right) 
\cdot \boldsymbol{\sigma}_{lr} + \boldsymbol{\mu}_{lr}
\end{equation}
where \(\mathrm{sr\_output}\) denotes the SR output, \(\boldsymbol{\mu}_{sr}\) and \(\boldsymbol{\sigma}_{sr}\) are the mean and standard deviation of \(\mathrm{sr\_output}\), and \(\boldsymbol{\mu}_{lr}\) and \(\boldsymbol{\sigma}_{lr}\) are those of the LR input.
Hence, the overall color distribution is aligned with the LR reference, mitigating color distortion to a considerable extent during the decoding process.

\begin{table*}[t]
    \centering
    \small
    \caption{Quantitative comparison on compressed video of Vid4 for 4× VSR. The PSNR (dB) and SSIM are calculated on Y-channel.
Red text indicates the best and blue text indicates the second best performance. The runtime is calculated by averaging the processing time across all images in the Vid4 dataset with CRF distortions, after testing on all subsets of the dataset.
 These model are carefully trained using the provided code.}
\resizebox{0.9\textwidth}{!}{
    \begin{tabular}{lcc|cccc|ccc}
        \toprule
        \multirow{2}{*}{Method} & \multirow{2}{*}{Params } & \multirow{2}{*}{Runtime}& \multicolumn{4}{c|}{Per clip with Compression CRF25} & \multicolumn{3}{c}{Average of clips with Compression} \\
        \cmidrule(lr){4-7} \cmidrule(lr){8-10}
        &(M) & (ms)& calendar & city & foliage & walk & CRF15 & CRF25 & CRF35 \\
        \midrule
 EDVR \cite{wang2019edvr} & 20.6 & 328& 21.69/0.672& 25.82/0.644& 22.01/0.559& 27.27/0.799& 25.95/0.794& 24.51/0.677&22.09/0.539\\
 BasicVSR \cite{chan2021basicvsr}& 5.9& 67& 22.06/0.687& 25.82/0.655& 22.62/0.577& 27.00/0.799& 26.03/0.795& 24.37/0.679&22.23/0.526\\
        IconVSR \cite{chan2021basicvsr} & 8.7 & 85& 21.36/0.634& 25.45/0.624& 23.89/0.597& 25.58/0.790& 25.79/0.739& 24.32/0.661& 22.27/0.529\\
        BasicVSR++ \cite{chan2022basicvsr++} & 7.3 & 99& 22.00/0.672& 25.90/0.658 & 23.91/0.609 & 27.23/0.804& 26.39/0.802& 24.71/0.689& 22.29/0.529\\
        RealBasicVSR \cite{chan2022investigating} & 6.3 & 77& 22.05/\textcolor{blue}{0.715}& \textcolor{blue}{26.01}/0.685& 23.30/0.602& 26.91/\textcolor{blue}{0.813}& 26.34/0.803& 24.57/\textcolor{blue}{0.713}& \textcolor{blue}{22.55}/\textcolor{red}{0.568}\\
        COMISR \cite{li2021comisr} & 6.2 & 88& 21.99/0.695& 26.00/\textcolor{blue}{0.689}& 23.44/0.612& 26.80/0.803& 26.44/0.805& 24.67/0.709& 22.35/0.544\\
        FTVSR  \cite{qiu2022learning} & 10.8 & 428& \textcolor{red}{22.44/0.730}& \textcolor{red}{26.34/0.700}& 23.94/\textcolor{red}{0.632}& \textcolor{red}{27.81/0.831}& 26.95/\textcolor{red}{0.821}& \textcolor{red}{25.24/0.728}& \textcolor{red}{22.66}/\textcolor{blue}{0.565}\\
        \midrule
 STDF + BasicVSR& 7.0
& 78& 21.88/0.684& 25.86/0.656& 23.50/0.606& 27.09/0.799& 26.44/0.808& 24.58/0.686&22.27/0.528\\
 STDF + BasicVSR++& 8.4
& 119& 22.01/0.672& 25.93/0.660& 23.88/0.609& \textcolor{blue}{27.33}/0.809& 26.49/0.812& 24.73/0.681&22.28/0.528\\
 STDF + FTVSR & 11.9& 458& 22.01/0.671& 25.81/0.650& 24.19/0.619& 26.77/0.777& 26.81/0.801& 24.71/0.686&22.28/0.529\\
        \midrule
        \rowcolor{blue!5} 
 FCA² + BasicVSR& 9.2& 33& 22.05/0.675& 25.78/0.650& 24.14/0.617& 26.97/0.798& 26.78/0.804& 24.74/0.685&22.29/0.529\\
 \rowcolor{blue!5} 
 FCA² + BasicVSR++& 10.6& 57& \textcolor{blue}{22.14}/0.678& 25.81/0.651& \textcolor{red}{24.19}/\textcolor{blue}{0.621}& 26.99/0.798& \textcolor{red}{27.01}/\textcolor{blue}{0.813}& \textcolor{blue}{24.78}/0.687&22.29/0.529\\
 \rowcolor{blue!5} 
 FCA² + FTVSR& 14.1& 111& 22.08/0.675& 25.80/0.651& 24.17/0.618& 26.97/0.797& 26.86/0.808& 24.75/0.685&22.28/0.529\\
        \bottomrule
    \end{tabular}
    }
    \label{quantitative_result}
    \vspace{-3mm}
\end{table*}

\section{Experiments}

\subsection{Implementation Details}
\subsubsection{\textbf{Compressed Datasets}} 
Following prior works \cite{li2021comisr,qiu2022learning,wang2023compression}, we first train our model on the Vimeo90k \cite{xue2019video} dataset, which comprises 64,612 video sequences (each consisting of seven frames at a resolution of \(448 \times 256\)), and subsequently evaluate on the Vid4 \cite{liu2013bayesian} dataset, containing four sequences of 30–50 frames each.
To further validate our approach, we also conduct experiments on the REDS \cite{nah2019ntire} dataset, following the methodology in \cite{qiu2022learning}. Compared to Vimeo90k, REDS features higher-resolution and higher-frame-rate videos, comprising 270 sequences of 100 frames each at a resolution of \(1280 \times 720\). This dataset encompasses a broader range of content types, allowing us to comprehensively assess our model's robustness and generalizability. For a fair comparison, we follow prior works \cite{chan2021basicvsr, wang2019edvr, li2020mucan} and evaluate on four sequences from the REDS dataset, commonly referred to as REDS4.

We follow the data pre-processing methodology established in prior work \cite{li2021comisr,qiu2022learning,wang2023compression}, applying identical operations and hyperparameter settings to ensure consistency and facilitate a fair comparison.
To synthesize LR frames and emulate compression artifacts, we first apply Gaussian smoothing (standard deviation = 1.5) to the HR frames and downsample by a factor of 4. We then compress the videos using the H.264 encoder \cite{wiegand2003overview} in FFmpeg 4.3, varying the Constant Rate Factor (CRF) among 15, 25, and 35. To simulate more realistic distortions, we further upload the encoded Vid4 videos to YouTube, download the compressed outputs, and use these distorted sequences for SR evaluation.
The performance of the proposed method is evaluated using Peak Signal-to-Noise Ratio (PSNR) and Structural Similarity Index (SSIM) \cite{wang2004image}, computed on the Y channel in the YCbCr color space.

\subsubsection{\textbf{Training Setting}} 
In FCA² training, we adopt a Cosine Annealing \cite{loshchilov2016sgdr} learning rate schedule and use the Adam \cite{kingma2014adam} optimizer with \(\beta_1 = 0.9\) and \(\beta_2 = 0.99\). The initial learning rate is set to \(1 \times 10^{-4}\). We keep the parameters of the optical flow network frozen for the first 5,000 iterations, after which they are unfrozen for subsequent training. 
For hyperparameters, the cleaning network is executed 3 times per forward pass, and we configure 9 channels with an overlap of 3 in the grouping module. We optimize the model using the CharbonnierLoss \cite{charbonnier1994two} function for 400k total iterations.

When integrating FCA² with other VSR networks, we strictly adhere to the parameter settings outlined in the original publications, ensuring consistency whether or not our module is incorporated. By maintaining identical configurations before and after integration, we underscore the migratory and pervasive nature of our approach. Additionally, we follow the training setup from previous work, using a dataset composed of 50\% uncompressed and 50\% compressed videos, where the compressed portions are uniformly sampled from diverse compression rates.
All models are trained for 400k iterations on A6000 servers using PyTorch. We make every effort to replicate the reported performance of each baseline model by rigorously matching the experimental conditions specified in their respective papers, ensuring consistency in all settings both before and after integration.

\begin{table*}[t]
    \centering
    \small
    \caption{Quantitative comparison on compressed video of REDS4 for 4× VSR. The PSNR (dB) and SSIM are calculated on Y-channel.
Red text indicates the best and blue text indicates the second best performance. The runtime is calculated by averaging the processing time across all images in the REDS4 dataset with CRF distortions, after testing on all subsets of the dataset. These model are carefully trained using the provided code.}
\resizebox{0.9\textwidth}{!}{
    \begin{tabular}{lcc|cccc|ccc}
        \toprule
        \multirow{2}{*}{Method}  & \multirow{2}{*}{Params } 
&\multirow{2}{*}{Runtime}
& \multicolumn{4}{c|}{Per clip with Compression CRF25} & \multicolumn{3}{c}{Average of clips with Compression} \\
        \cmidrule(lr){4-7} \cmidrule(lr){8-10}
         & (M) 
&(ms)
& Clip 000& Clip 011& Clip 015& Clip 020& CRF15 & CRF25 & CRF35 \\
        \midrule
        EDVR \cite{wang2019edvr} 
& 20.6 
&385& 24.38/0.629& 26.01/0.702& 28.30/0.783& 25.21/0.708& 28.72/0.805& 25.98/0.706& 23.36/0.600\\
 BasicVSR \cite{chan2021basicvsr}
& 5.9
&78& 24.37/0.628& 26.01/0.702& 28.13/0.777& 25.21/0.709& 29.05/0.814& 25.93/0.704&23.22/0.596\\
        IconVSR \cite{chan2021basicvsr} 
& 8.7 
&93& 24.35/0.627& 26.00/0.702& 28.16/0.777& 25.22/0.709& 29.10/0.816& 25.93/0.704& 23.22/0.596\\
        BasicVSR++ \cite{chan2022basicvsr++} 
& 7.3 
&108& 24.88/0.650& 26.11/0.710& 28.98/0.788& 25.25/0.710& 29.20/0.815& 25.96/0.705& 23.30/0.600\\
        RealBasicVSR \cite{chan2022investigating} 
& 6.3 
&80& 25.12/0.671& 26.88/0.730& 29.00/0.800& 25.88/0.752& 29.31/0.814& 26.55/0.730& 23.68/0.610\\
        COMISR \cite{li2021comisr} 
& 6.2 
&98& 24.76/0.660& 26.54/0.722& 29.14/0.805& 25.44/0.724& 28.40/0.809& 26.47/0.728& 23.56/0.599\\
        FTVSR  \cite{qiu2022learning} 
& 10.8 
&489& 26.06/\textcolor{red}{0.703}& \textcolor{red}{28.71}/\textcolor{blue}{0.779}& 30.17/0.839& \textcolor{red}{27.26}/\textcolor{red}{0.782}& \textcolor{red}{30.51}/\textcolor{red}{0.853}& 28.05/\textcolor{red}{0.776}& 24.82/0.657\\
        \midrule
 STDF + BasicVSR& 7.0
&86& 24.40/0.622& 26.11/0.705& 28.15/0.777& 25.30/0.710& 29.05/0.815& 25.96/0.705&23.28/0.599\\
 STDF + BasicVSR++ &

8.4&135& 25.00/0.660& 26.21/0.715& 29.01/0.805& 25.77/0.748& 29.25/0.815& 26.00/0.710&23.50/0.609\\
 STDF + FTVSR & 11.9&531& 25.88/\textcolor{blue}{0.695}& 28.25/0.751& 29.85/0.822& 26.58/0.750& 28.82/0.810& 27.71/0.750&24.48/0.640\\
        \midrule
        \rowcolor{blue!5} 
 FCA² + BasicVSR & 
9.2&37& 25.55/0.670& 27.11/0.740& 29.22/0.810& 26.35/0.758& 29.25/0.810& 27.00/0.748&24.14/0.622\\
\rowcolor{blue!5} 
 FCA² + BasicVSR++ & 10.6&59& \textcolor{red}{26.37}/0.687& \textcolor{blue}{28.47}/\textcolor{red}{0.780}& \textcolor{blue}{30.78}/\textcolor{red}{0.859}& \textcolor{blue}{27.22}/\textcolor{blue}{0.762}& \textcolor{blue}{29.58}/0.820& \textcolor{red}{28.19}/\textcolor{blue}{0.772}&\textcolor{blue}{25.70}/\textcolor{red}{0.693}\\
 \rowcolor{blue!5} 
 FCA² + FTVSR & 14.1&120& \textcolor{blue}{26.21}/0.685& 28.26/0.767& \textcolor{red}{31.02}/\textcolor{blue}{0.856}& 26.91/0.758& 29.44/\textcolor{blue}{0.831}& \textcolor{blue}{28.11}/0.767&\textcolor{red}{25.74}/\textcolor{blue}{0.690}\\
        \bottomrule
    \end{tabular}
    \label{quantitative_result2}
    }
\end{table*}

\begin{figure*}[t]
    \centering
    \includegraphics[width=0.85\linewidth]{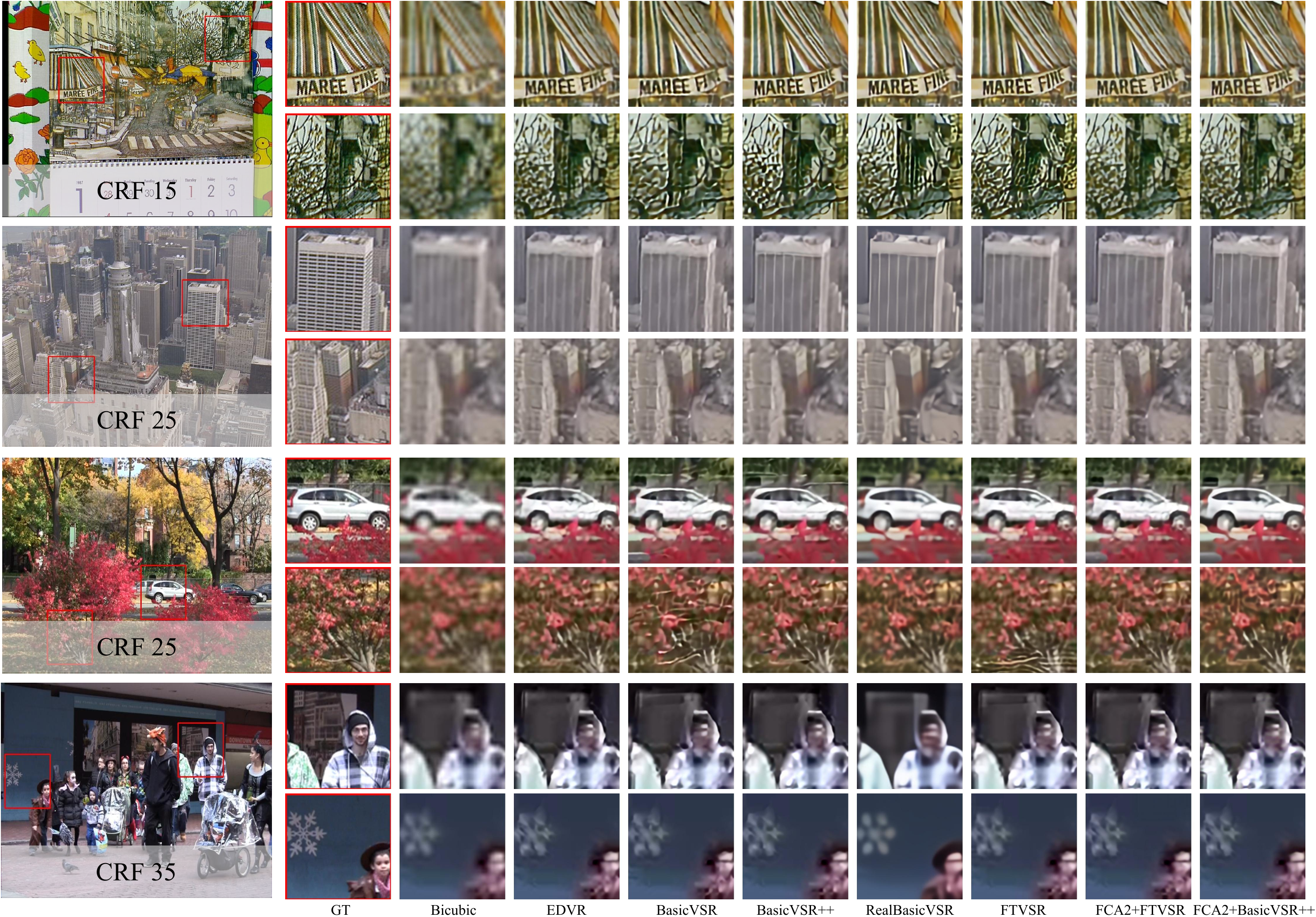}
    \caption{Qualitative comparison on the compressed Vid4 test set for 4× VSR. Zoom in for better visualization.}
    \label{Vid4}
    \vspace{-3mm}
\end{figure*}

\subsection{Comparison with State-of-the-Arts}

\subsubsection{\textbf{Quantitative Evaluation on Compressed Vid4 for 4× VSR}}
In this section, we compare several state-of-the-art models, including EDVR \cite{wang2019edvr}, BasicVSR \cite{chan2021basicvsr}, IconVSR \cite{chan2021basicvsr}, BasicVSR++ \cite{chan2022basicvsr++}, RealBasicVSR
\cite{chan2022investigating}, COMISR \cite{li2021comisr}, and FTVSR \cite{qiu2022learning}. The first four are VSR models designed for a single idealized distortion, RealBasicVSR addresses blind SR, and the latter two have been specifically developed for compression artifacts. Additionally, to demonstrate the superiority of our proposed method’s modular design, we also include a comparison with the modularly designed STDF \cite{deng2020spatio} approach.

We trained the aforementioned models on the Vimeo90k training set to faithfully reproduce their true performance. As shown in Table \ref{quantitative_result}, our model, designed in a modular fashion, can be seamlessly integrated with various existing models. With only a minimal increase in parameters, it significantly reduces inference time while enhancing performance. When combined with FTVSR, our model achieves nearly identical performance at a substantially lower inference time. Moreover, when integrated with models like BasicVSR++ that already leverage strong inter-frame information, our model further enhances this capability by compressing adjacent frames and eliminating redundant information, ultimately yielding improved results.
Although STDF can enhance model performance to some extent, it also increases both the parameter count and inference time. Overall, our modular design exhibits strong transferability, effectively reducing inference time while improving model performance. Notably, it achieves robust performance across various compression levels.

Additionally, we train and test our model on the REDS dataset, as shown in Table \ref{quantitative_result2}. The REDS dataset features higher resolution and more frames compared to the Vimeo90k dataset (100 fps for REDS and 7 fps for Vimeo90k). In the Vimeo90k dataset, with only 7 frames per second, the inter-frame differences are large, and frame compression does not significantly improve performance. However, when the video quality increases, as in the case of the REDS dataset, the inter-frame differences become smaller, and frame compression effectively removes redundant information. This enhances the utilization of inter-frame information and significantly boosts model performance. Moreover, the speedup in inference time is more evident in the SR tests for higher-quality videos, highlighting the advantages of our model.

\subsubsection{\textbf{Qualitative Comparison on Compressed Vid4 for 4× VSR}}
We also select several models for visual comparison, including conventional VSR methods, blind SR approaches, compression-aware SR techniques, and results obtained by integrating our modular design. We present outcomes corresponding to three different compression levels (CRF15, CRF25, and CRF35) to demonstrate the model’s generalizability. As shown in Fig. \ref{Vid4}, incorporating our method produces results that are overall smoother with sharper texture details. Notably, across various compression levels, the performance remains robust; for example, in the CRF35 case, our best method (combined with BasicVSR++) yields markedly clearer edges in the background snowflake details. In summary, our modular design can effectively improve model performance.

\begin{figure}[t]
    \centering
    \includegraphics[width=0.9\linewidth]{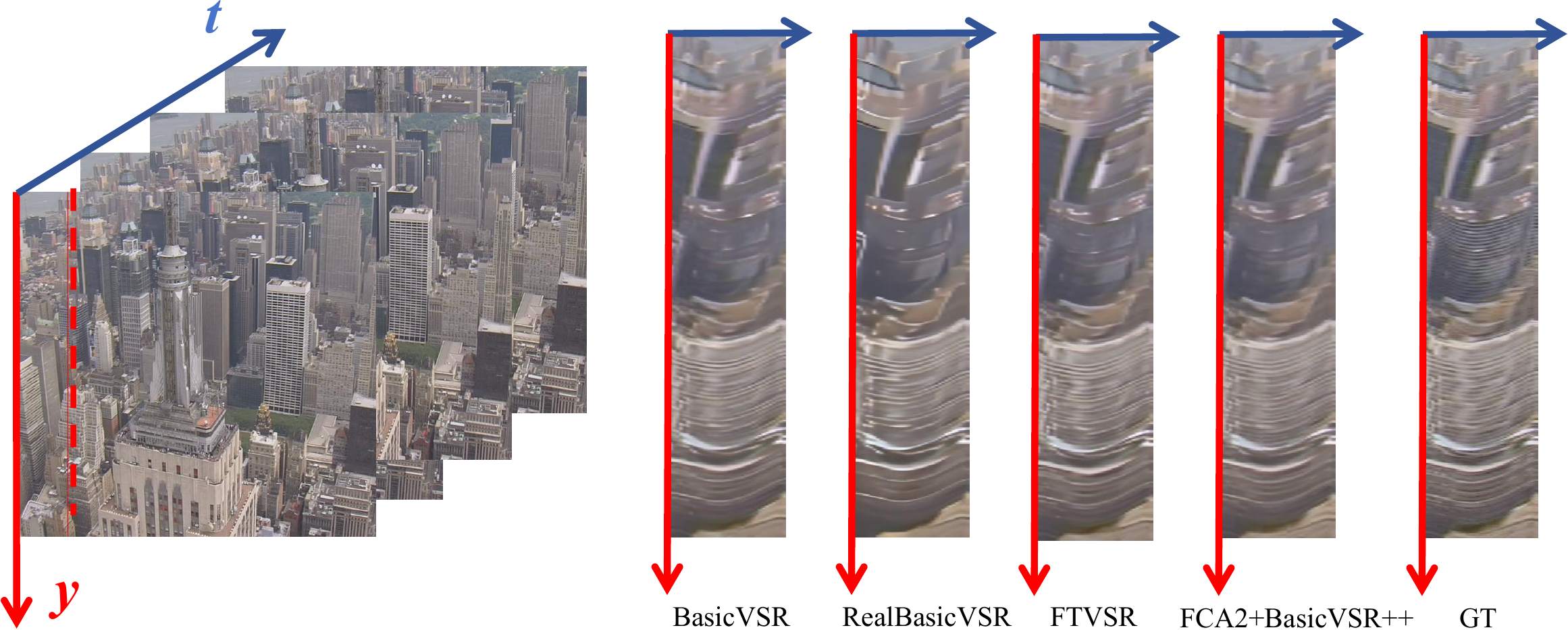}
    \caption{Qualitative comparison in the time dimension on the Vid4 dataset, with CRF compression level set to 25.}
    \label{time}
    \vspace{-3mm}
\end{figure}

\subsubsection{\textbf{Time-Domain Qualitative Comparison for 4× VSR on Compressed Vid4}}
To further validate the effectiveness of our model, we display the SR results along the temporal dimension by selecting a column from an image and showcasing its changes over time. We compare several representative methods. As shown in Fig. \ref{time}, our results exhibit smoother variations over time within the same spatial region, with reduced discontinuities between frames and a closer resemblance to the ground truth distribution. In summary, these findings confirm the superior temporal consistency and overall robustness of our approach.

\begin{figure*}[t]
    \centering
    \includegraphics[width=0.7\linewidth]{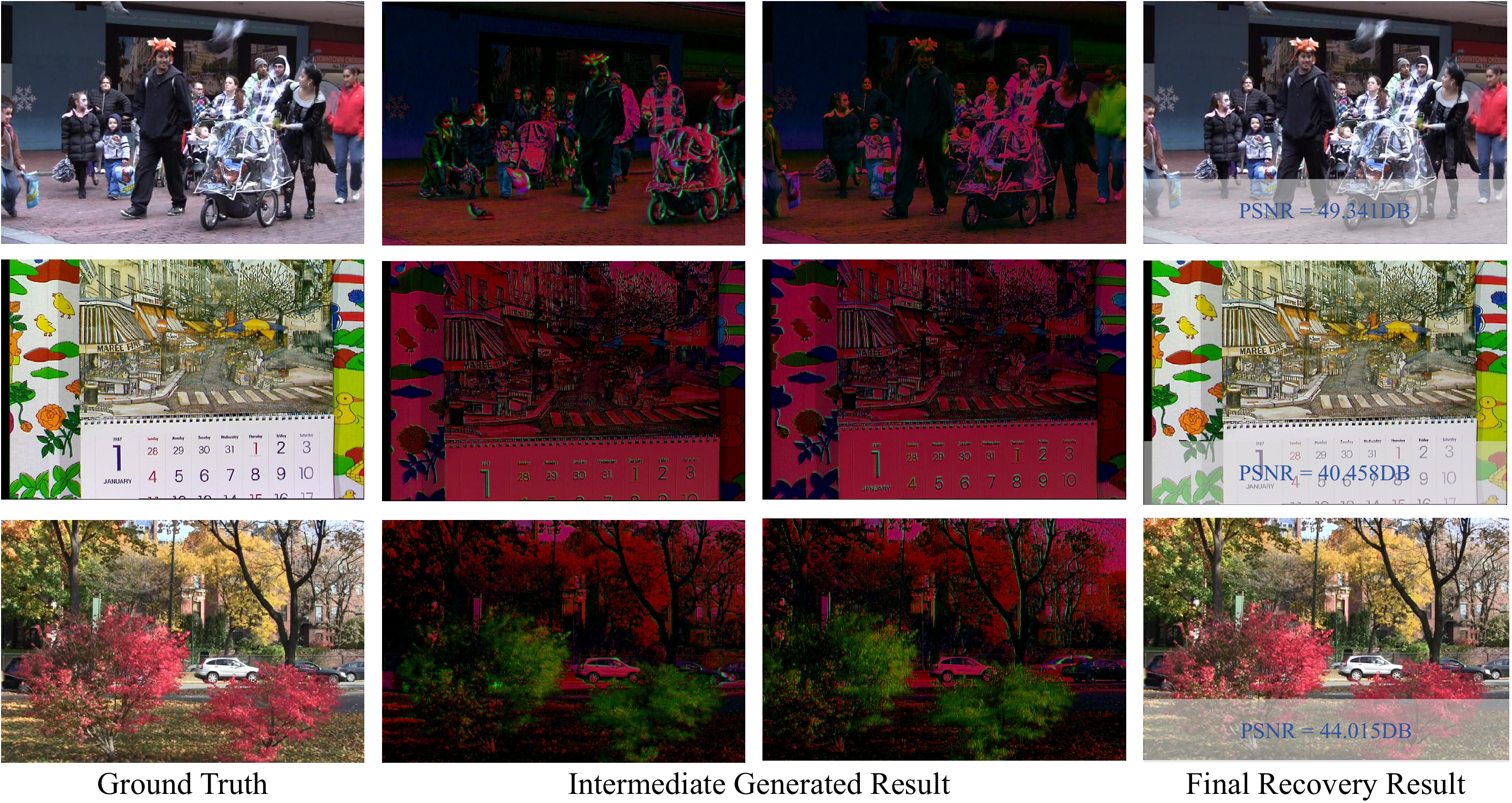}
    \caption{Visualization of FCA² encoding results on the compressed Vid4 dataset (CRF15). From left to right: the original image, the intermediate generated image, and the final recovered result. The intermediate results are obtained from the first and last frames after frame compression.}
    \label{mid_img}
\end{figure*}

\begin{figure*}[t]
    \centering
    \includegraphics[width=0.7\linewidth]{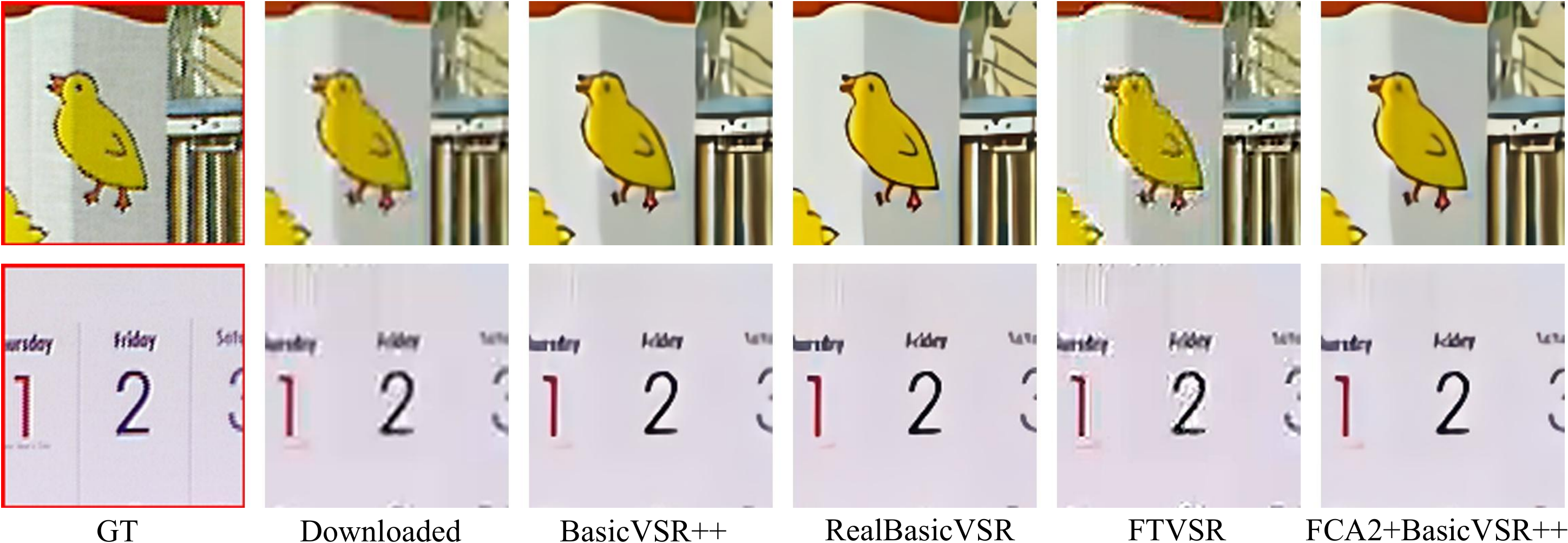}
    \caption{Qualitative comparison on Vid4 downloaded from
YouTube for × 4 VSR}
    \label{Youtube}
    \vspace{-3mm}
\end{figure*}

\subsubsection{\textbf{Visualization of Intermediate FCA² Encoder Representations on Compressed Vid4}}
To further validate the effectiveness of our model’s encoding-decoding process, we visualize the intermediate compressed encoding frames alongside the final decoded outputs, as shown in Fig. \ref{mid_img}. 
It can be seen the proposed compression and decoding framework achieves visually consistent reconstruction with the original images, preserving structural integrity and perceptually critical details. Intermediate feature representations maintain precise spatial alignment with source content, retaining human-visually recognizable textures and edges while ensuring high-fidelity visual output.

The overlapping strategy, coupled with the incorporation of optical flow information, leads to noticeable color distortion to some extent.
While the method demonstrates robust detail preservation in static regions—where images exhibit sharp edges, clear texture hierarchies, and enhanced resolution—certain dynamic areas may present subtle artifacts. These artifacts primarily stem from motion blur during compression, manifesting as localized blurring or mild distortions in high-motion regions. This phenomenon aligns with inherent challenges in temporal consistency modeling for dynamic content compression.
The framework effectively balances compression efficiency with perceptual quality, particularly excelling in static scene reconstruction through spatially optimized feature encoding. By strategically allocating computational resources based on motion characteristics, it maintains high visual fidelity across stationary regions while mitigating artifacts in dynamic portions through adaptive temporal regularization. This approach ensures efficient data compression without compromising human-visually salient details, establishing a practical trade-off between computational overhead and reconstruction accuracy.

\subsection{Experiments on Real-world Compressed Videos}
To further highlight the compression-aware capabilities of our model, we conduct experiments on real-world distorted videos following the procedure in COMISR \cite{li2021comisr} and FTVSR \cite{qiu2022learning}. Specifically, we first upload raw videos to YouTube, allowing the platform’s built-in pipeline to encode and compress them, and then download the compressed versions for SR. As illustrated in Fig. \ref{Youtube}, magnified views reveal that our model restores fine details more accurately than competing approaches, yielding clearer outlines for letters, numbers, and small objects such as chicks. These results demonstrate the effectiveness of our model in tackling real-world compression artifacts.


\subsection{Comparative Inference Speed across Diverse Frame Rates and Resolutions}

\begin{figure*}
    \centering
    \includegraphics[width=0.85\linewidth]{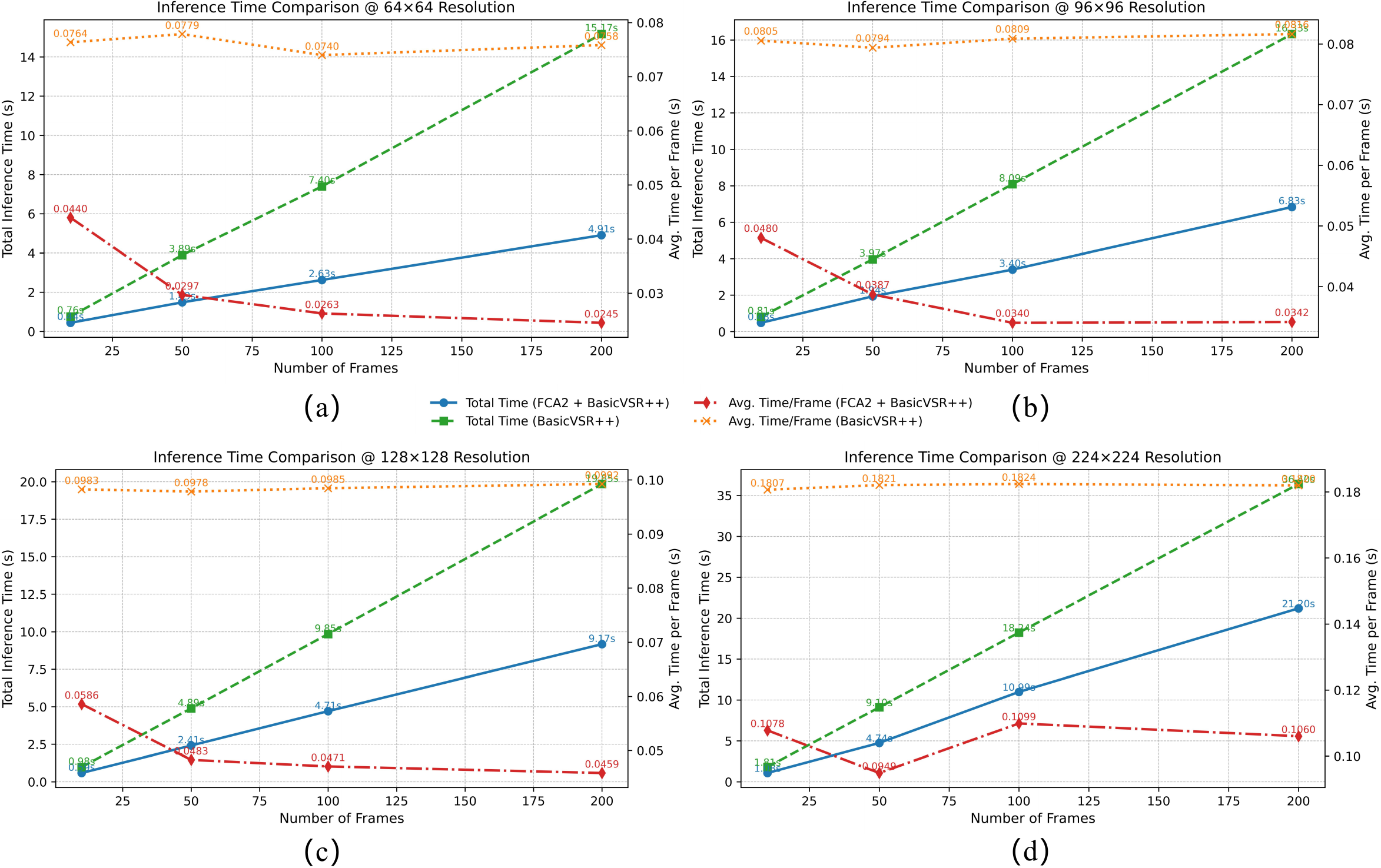}
    \caption{The inference time of the proposed model under varying frame rates and resolutions for super-resolution tasks. Subfigures (a)–(d) illustrate a comparative analysis between our method and the non-frame-compression-based VSR model, BasicVSR++, across different video resolutions and frame rates.}
    \label{time_cost}
\end{figure*}

To further demonstrate the efficiency improvements of our method in terms of inference speed, we conduct extensive simulations across a diverse set of video scenarios. Specifically, we evaluate videos with varying frame counts—10, 50, 100, and 200 frames—progressively covering short to long sequences. In addition, we consider different video resolutions ranging from $64 \times 64$ to $224 \times 224$. The detailed inference results are illustrated in Fig. \ref{time_cost}.
Firstly, it can be observed that the proposed frame compression strategy consistently accelerates inference across all tested scenarios. Along the frame rate dimension, the results reveal that as the number of frames increases, the proportion of frames being compressed also increases, leading to more substantial speedups. In contrast, the inference speed of non-frame-compressed VSR methods remains relatively insensitive to the number of frames. Furthermore, when analyzing across different resolutions, we observe that higher resolutions naturally incur longer inference times; however, our method continues to offer significant efficiency gains, demonstrating its robustness and scalability.

\begin{table}[t]
  \caption{Ablation Study on FCA² Modules. Results Are Tested on Vid4 Dataset.}
    \centering
 \resizebox{0.45\textwidth}{!}{
    \begin{tabular}{ccc|cc}
    \toprule
         Group Encoding&  Flow Guidance&  Color Restoration&  PSNR $\uparrow$& SSIM $\uparrow$ \\
         \midrule
         &  &  &  27.84& 0.870\\
         \checkmark&  &  &  37.41& 0.974\\
         \checkmark&  \checkmark&  &  42.79& 0.994\\
         \checkmark&  &  \checkmark&  39.79& 0.982\\
         \rowcolor{blue!5} 
 \checkmark& \checkmark& \checkmark& 45.86&0.996\\
 \bottomrule
    \end{tabular}
    }
  
    \label{abl1}
\end{table}

\begin{table}[t]
  \caption{Ablation Study on the encoder of FCA² Modules. Results Are Tested on Vid4 Dataset.}
    \centering
    \small
 \resizebox{0.45\textwidth}{!}{
    \begin{tabular}{cc|ccc}
    \toprule
           Frame Encoding&  CRF Cleaning&  PSNR $\uparrow$& SSIM $\uparrow$  &Runing Time (ms) $\downarrow$\\
           \midrule
           &  &  26.39&  0.802&99\\
           \checkmark&  &  26.79&  0.805&50\\
           &  \checkmark&  26.58&  0.808&106\\
           \rowcolor{blue!5} 
  \checkmark& \checkmark& 27.01& 0.813&57\\
 \bottomrule
    \end{tabular}
    }
  
    \label{abl2}
    \vspace{-3mm}
\end{table}

\subsection{Ablation Studies:}

\subsubsection{\textbf{On the individual components’ functionalities}}
In this section, we evaluate the effectiveness of the key modules in our FCA² model on the Vid4 dataset. Specifically, we analyze the impact of three main components: the grouped coding strategy, the optical flow-guided coding strategy, and the color correction module, as summarized in Table \ref{abl1}.
Our results indicate that the grouped coding strategy plays a crucial role in maintaining reconstruction quality. Directly compressing multiple frames into a fixed number in a single step results in substantial feature loss, leading to a PSNR of only 27.84 dB. In contrast, grouping a subset of frames before compression significantly improves performance. Additionally, incorporating optical flow guidance and performing color correction further enhance coding accuracy. Optical flow guidance, a widely adopted approach in VSR, helps the model capture dynamic spatial relationships between frames. Meanwhile, due to the nature of our encoding process, which selectively encodes multiple color channels, distortions in color distribution may arise, as illustrated in Fig. \ref{mid_img}. Applying color correction effectively mitigates these distortions, further boosting overall performance.
In summary, the proposed modules complement each other, collectively contributing to the best reconstruction quality.

\begin{table}[t]
  \caption{Ablation Study on the Training Mode of FCA². Results Are Tested on Vid4 Dataset.}
    \centering
    \small
 \resizebox{0.45\textwidth}{!}{
    \begin{tabular}{c|ccc}
    \toprule
           Training Mode&  CRF15 & CRF25 &CRF35 \\
           \midrule
           \rowcolor{blue!5} 
           Pre-training&  27.01/0.813&  24.78/0.687&22.29/0.529\\
           Joint Training&  25.70/0.757&  23.79/0.668&22.25/0.529\\
           Pre-training + Joint Training&  26.86/0.808&  24.75/0.685&22.28/0.529\\
    \bottomrule
    \end{tabular}
    }
    \vspace{-3mm}
    \label{abl3}
\end{table}

\begin{table}[t]
    \caption{Ablation results for different settings of the number of subgroups bands ($n_{subs}$) and the number of overlaps bands ($n_{ovls}$) on the Vid4 dataset at scale 4.
    }
    \centering
    \small
 \resizebox{0.45\textwidth}{!}{
    \begin{tabular}{c |c c c}
    \toprule
    Settings  & PSNR $\uparrow$& SSIM $\uparrow$ &Runing Time (ms) $\downarrow$\\
    \midrule
    $n_{subs} = 6, n_{ovls} = 0$&46.15&0.999&44\\
    $n_{subs} = 6, n_{ovls} = 2$&47.99&0.999&45\\
    $n_{subs} = 9, n_{ovls} = 0$&43.75&0.995&30\\
    \rowcolor{blue!5} 
    $n_{subs} = 9, n_{ovls} = 3$&45.86&0.996&33\\
    $n_{subs} = 12, n_{ovls} = 4$&37.24&0.974&29\\
    $n_{subs} = 18, n_{ovls} = 8$&33.11&0.966&22\\
    \bottomrule
    \end{tabular}
    }
    \label{group}
    \vspace{-3mm}
\end{table}

\subsubsection{\textbf{On the encoder of FCA² Modules}}
In this section, we assess the effectiveness of the coding module in FCA² for compression-aware SR, which consists of two processes: frame compression and CRF cleaning. To validate their impact, we design experiments where FCA² is integrated with BasicVSR++, as detailed in Table \ref{abl2}. Our results indicate that both processes contribute to performance enhancement. Specifically, frame compression significantly reduces inference time and improves the utilization of inter-frame information, while the combined application of frame compression and CRF cleaning yields the best overall results.


  

\subsubsection{\textbf{On the Training Mode of FCA²}}
In our proposed framework, the FCA² module is initially pre-trained to develop robust image encoding capabilities before being integrated with the VSR network, which is subsequently trained to enhance the encoded intermediate features. To investigate the effectiveness of different training strategies, we design ablation experiments comparing three approaches.
As shown in Table \ref{abl3}, jointly training the entire model from scratch leads to suboptimal performance. This is primarily due to the additional parameters introduced by FCA², which results in a considerably larger model that struggles to converge when all parameters are updated simultaneously. Notably, FCA² is intended to learn a near lossless compression and decompression of images, rather than directly performing SR, and unified training fails to satisfy this specialized requirement. In contrast, fixing the pre-trained FCA² parameters during VSR training achieves superior performance. Although allowing FCA²’s parameters to be fine-tuned during subsequent training slightly degrades its coding and decoding effectiveness, the negative impact is minimal. Overall, the best results are obtained by using pre-trained and fixed FCA² parameters, thereby ensuring a lossless encoding-decoding process that optimally supports the SR task.

\begin{table}[t]
  \caption{Ablation Study on the Placement of the Color Correction Module. Results Are Tested on Vid4 Dataset.}
    \centering
    \begin{tabular}{cc|cc}
    \toprule
           Encoder&  Decoder&  PSNR $\uparrow$& SSIM $\uparrow$  \\
           \midrule
           &  &  42.79&  0.994\\
           \checkmark&  &  37.51&  0.966\\
           \rowcolor{blue!5} 
           &  \checkmark&  45.86&  0.996\\
  \checkmark& \checkmark& 40.57& 0.972\\
 \bottomrule
    \end{tabular}
    \vspace{-3mm}
    \label{abl4}
\end{table}

\begin{table}[t]
  \caption{Cross-Dataset Evaluation of Video Frame Compression Performance.}
    \centering
    \begin{tabular}{cc|cc}
    \toprule
           Training&Testing& PSNR $\uparrow$&SSIM $\uparrow$ \\
           \midrule
            Vimeo90k&REDS4&  45.52&0.995\\
           REDS&Vid4&  44.17&0.994\\
    \bottomrule
    \end{tabular}
  
    \label{abl5}
    \vspace{-3mm}
\end{table}

\subsubsection{\textbf{On the different settings of the number of subgroups bands}}
When applying our model, one must specify both the number of channels to be compressed and the number of overlapping channels, as summarized in Table \ref{group}. Incorporating this overlap strategy improves overall performance by enhancing channel-wise information utilization, consistent with similar techniques in prior work \cite{jiang2020learning, wang2024enhancing, wang2022group}. Additionally, increasing the number of simultaneously compressed channels yields a slight decrease in accuracy but provides faster inference. Our chosen parameters represent a compromise between these two metrics.



  

\subsubsection{\textbf{On the Placement of the Color Correction Module}}
To further improve the model's encoding and decoding capabilities and mitigate color distortion issues in multi-frame encoding, we incorporate a dedicated color correction module. We empirically investigate different placements of the module—during encoding, decoding, or both—and summarize the results in Table \ref{abl4}.
As demonstrated, the inclusion of color correction substantially enhances the encoder-decoder performance, striving for minimal information loss in frame compression and recovery. However, applying color correction at the encoding stage must be approached with caution. Since the module adjusts the three RGB channels by fitting them to typical single-frame color distributions, enforcing this transformation on multi-frame representations (which do not strictly map to RGB) may introduce inaccuracies. Our experiments confirm this limitation, and therefore, color correction is exclusively applied at the final decoding stage to refine the output without disrupting the encoding process.

\subsubsection{\textbf{On the Cross-Dataset Evaluation of Video Frame Compression Performance}}
Our model is initially pre-trained on a large-scale dataset to establish robust encoding and decoding capabilities, effectively learning HR feature distributions for frame compression and reconstruction. To validate its generalization and confirm that it does not overfit to a specific dataset, we conduct cross-dataset training experiments. Specifically, we train on Vimeo90k and evaluate on REDS4, as well as train on REDS and evaluate on Vid4, as reported in Table \ref{abl5}. The results indicate that the proposed model does not depend on a particular dataset; rather, it achieves strong encoding and decoding performance across different datasets once it has been trained on a suitable HR dataset.

\section{Limitation and Future Work}
Although our approach effectively improves inference speed by enhancing inter-frame information utilization, several limitations remain. First, the frame compression module requires pre-training, and the overall framework is not yet a fully end-to-end one-stage model, which may hinder practical deployment efficiency. Second, there is still room for optimization in both the compression and fusion strategies across frames. Specifically, for low frame rate videos—where the temporal difference between frames is significant—fewer frames should be compressed to preserve critical temporal cues. In contrast, for high-quality videos with high frame rates, greater compression can be applied to reduce redundancy. Therefore, designing an adaptive frame compression mechanism based on frame rate and video content characteristics could further enhance both model performance and practical applicability.
In future work, we aim to address the aforementioned limitations and further extend our approach to other video-related tasks, such as video quality assessment and beyond.

\section{Conclusion}
In this paper, we propose a frame-aware compression module for CVSR that enhances inter-frame information utilization and significantly accelerates inference. Specifically, we design a frame compression coding strategy for video sequences by integrating channel compression coding techniques from HSI-SR. To achieve efficient frame compression and decompression, we incorporate an optical flow-guided coding module, a frame-based group coding strategy, and a color correction module. Our module demonstrates strong adaptability, allowing it to be seamlessly integrated with most existing VSR frameworks to improve inter-frame information utilization and accelerate inference speed. Experimental results validate the effectiveness of our approach, achieving near SOTA performance. Furthermore, this frame compression approach offers a new research direction for video enhancement tasks within low-level vision domains. In conclusion, our method effectively addresses key challenges in CVSR and provides novel insights for related fields.

\bibliographystyle{IEEEtran}
\bibliography{reference}

\end{document}